\documentclass[conference]{IEEEtran}
\usepackage{graphicx}
\usepackage[cmex10]{amsmath}
\usepackage{algorithmic}
\usepackage{array}
\usepackage[tight,footnotesize]{subfigure}
\usepackage{fixltx2e}
\usepackage{url}

\begin{document}

\title{Reconciliation for Practical Quantum Key Distribution with BB84 protocol}

\author{\IEEEauthorblockN{N. Benletaief, H. Rezig *Members IEEE, A. Bouallegue *Members IEEE}
\IEEEauthorblockA{Communication System laboratory Sys'Com \\National Engineering School of Tunis\\ BP 37, 1002 Tunis  Belvédère, Tunisia\\
Emails: benletaief.nedra@gmail.com, houria.rezig@enit.rnu.tn, ammar.bouallegue@enit.rnu.tn}}

\maketitle

\begin{abstract}

This paper investigates a new information reconciliation method for quantum key distribution in the case where two parties exchange key in the presence of a malevolent eavesdropper. We have observed that reconciliation is a special case of channel coding and for that existing techniques can be adapted for reconciliation. We describe an explicit reconciliation method based on Turbo codes. We believe that the proposed method can improve the efficiency of quantum key distribution protocols based on discrete quantum states.

\end{abstract}


\section{Introduction}

Cryptography is the art of hiding information in a string of bits meaningless to any unauthorized party. To achieve this goal, a message is combined according to an algorithm with some additional secret information which is the key to produce a cryptogram. In the traditional terminology, Alice is the party encrypting and transmitting the message, Bob the one receiving it, and Eve the malevolent eavesdropper.
For a crypto-system to be considered secure, it should be impossible to unlock the cryptogram without owning the key.\\ At this point, quantum key distribution enters the scene by allowing two physically separated parties to create a random secret key and to verify that the key has not been intercepted. The idea is to use bits encoded on individual photons to exchange cryptographic keys between network users where the security is based on the laws of physics rather than in computational complexity (as is the case for the most classical cryptographic approaches).\\
This manuscript is organized as follows: In Section 2, we adress the problem statement. In Section 3, we present an overview of related works. In Section 4, we describe our proposed reconciliation method and some conclusions
and perspectives are consequently drawn in Section 5.

\section{Problem statement: BB84 protocol}
 The first scheme for quantum cryptography was introduced by Bennett and Brassard~\cite{Bennett} in 1984. But, the first experimental demonstration of the protocol was performed in 1991 using the polarization states of single
photons to transmit a random key. A basis is chosen to distinguish the two values 0 and 1 without ambiguity. One choice is the rectilinear basis $\bigoplus$ where photons are polarized at angle $0^{\circ}$ ($\leftrightarrow$) or $90^{\circ}$ ($\updownarrow$) representing 0 and 1 respectively. Another choice is the diagonal basis $\bigotimes$ where 0 is represented by photons polarized at $45^{\circ}$ ($\nearrow$) and 1 by photons polarized at $135^{\circ}$ ($\nwarrow$). \\Protocol BB84 like any other protocol of quantum cryptography is based on two principal phases: a quantum phase via a one-way physical quantum channel and a public phase using an
authenticated two-way classic ideal channel. The four steps of these phases will be illustrated below.

\begin{figure}[h]
    \centering
    \includegraphics[width=9cm,height=8cm]{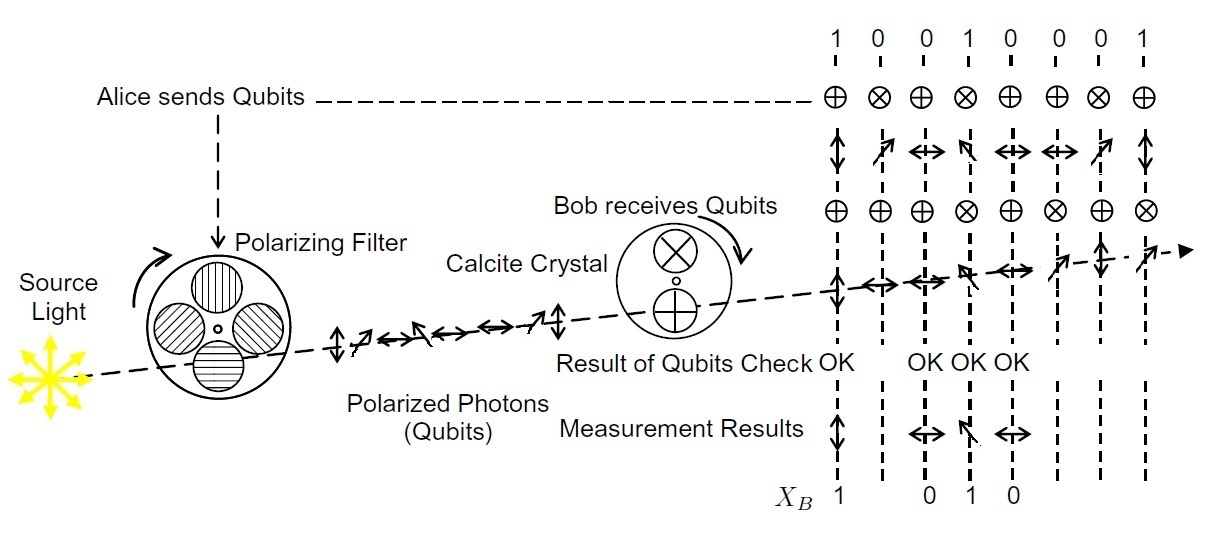}

    \caption{First step of BB84 protocol~\cite{fig}.\label{Fig:BB84}}
\end{figure}

First, Alice sends quantum states to Bob on the quantum channel and Bob measures these states. On receiving the state, Bob informs her via public channels of the basis used to accept each bit.  Alice informs Bob, which bits were applied the correct basis.  Then,  they discard the incorrect ones and use the remaining. This process gives the two parties correlated random variables, called sifted key $X_{A}$ and $X_{B}$ (Fig.~\ref{Fig:BB84}).

In order to detect eavesdropping, Alice and Bob compare a sample of the transmitted data throw a classical public authenticated channel and thus can determine an upper bound on the amount of information a possible eavesdropper may have acquired. \\Then, a step called reconciliation is done. Alice and Bob exchange information over the public authenticated channel in such a way that Bob can recover $X_{A}$ knowing $X_{B}$. The exchanged information is considered known to an eavesdropper. Finally, another step is essential and it consists in applying a privacy amplification protocol to wipe out the enemy's information on both quantum and classical transmissions, at the cost of a reduction in the key length.
\\It thus appears clearly that reconciliation is crucial and tricky (we should try to decrease the amount of processus exchanged information). Hence, the interest of investigating the use of an efficient reconciliation method in this context. So, in the next section, we present reported solutions in the literature to tackle such problem.

\section{Related works}
Several methods of error reconciliation for quantum cryptography have been already reported in the literature.  In what follows, we will only focus on the most employed ones, such as the best known Cascade algorithm~\cite{Cascade}, the Binary algorithm~\cite{Binary}, the Winnow algorithm~\cite{Winnow}, and the Low-Density Parity-Check code LDPC~\cite{LDPC}.\\ Each method have some positive and negative aspects. For exemple, Cascade and Binary removes a single error and don't introduce additional errors to multiple errors block, instead of the Winnow algorithm because the Hamming algorithm only reveals one single error in each block.
Also, the Cascade and the Binary require significant more time of communication that it is proportional to the length of key. While for the winnow algorithm, communication time only depends on the error rate. Generally, compared to Cascade method, the LDPC codes can correct the same range of errors but has the advantage to improve the safety of the used protocol. This has motivated our approach to study turbo codes which had been proved to perform better when compared with LDPC in the literature. As far as we are concerned by reconciliation method based on turbo codes, we will describe it in more details in the next section.

\section{ Proposed reconciliation Quantum Key distribution method}
Quantum cryptography asserts that the intervention of an eavesdropper inevitably introduces
transmission errors and, therefore can be detected by legitimate parties.
That is why reconciliation or Error Correction will be necessary for preserving the key against not only eavesdropping, but also channel noise and other unwanted interactions in quantum computation and communication. \\In our study, we limit our researches to the presence of a particular type of eavesdropping attack: intercept and resend. We develop a reconciliation method based on the well known turbo codes.
\\We start with a presentation of the turbo code theoretic tool and a brief explanation of intercept and resend attack principle. Then, we move to the description of the method and finally we conclude from experiments.

\subsection{Turbo codes principle}
\begin{figure}[h]
    \centering
    \includegraphics[width=9cm,height=5.5cm]{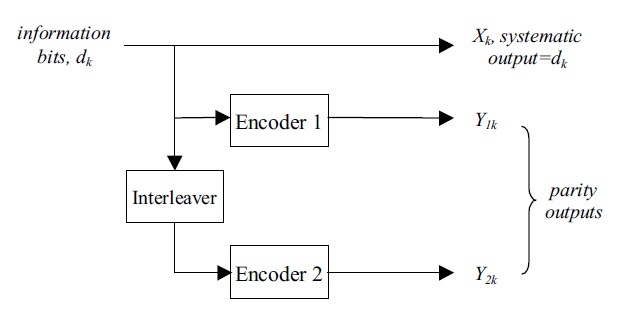}
 \caption{The generic turbo encoder.\label{Fig:turbo}}
\end{figure}

\begin{figure}[h]
    \centering
    \includegraphics[width=9cm,height=5.5cm]{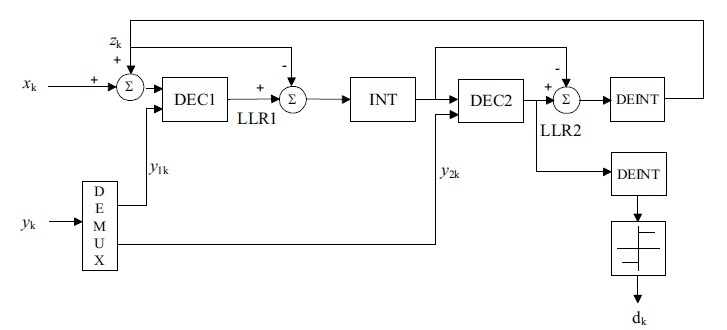}
 \caption{The generic turbo decoder.\label{Fig:turbodec}}
\end{figure}

Turbo coding was first introduced in 1993 by Berrou and $al.$~\cite{turbo}. It consist on a parallel concatenation of two, or more, constituent codes separated by one, or more, interleavers (Fig.~\ref{Fig:turbo}). The constituent codes are usually two identical recursive systematic convolutional codes. The input sequence to be encoded is divided into blocks of length $N$. Each block is encoded by the first encoder and interleaved before passing through the second encoder. Researches have proved that turbo codes can approach the Shannon limit closer than any other known forward error correcting code. The efficiency of the turbo codes is due to the use of an iterative process at the decoder side and the presence of an interleaver at the encoder side, which adds randomness-like effect to the code.
\\The turbo decoder consists of two, or more, Soft-In Soft-Out (SISO) maximum likelihood decoders (Fig.~\ref{Fig:turbodec}).

\begin{figure}[!htbp]
    \centering
    \includegraphics[width=8cm,height=3cm]{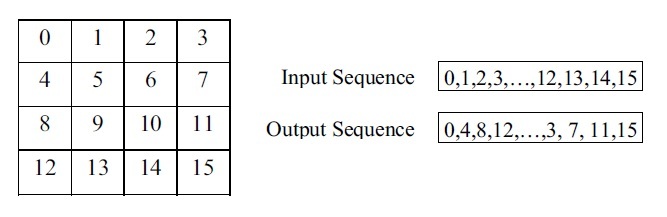}
 \caption{Row-column interleaver.\label{Fig:interleaver}}
\end{figure}

Decoders are operating in parallel, exchanging iteratively extrinsic information. Two families of decoding algorithms are commonly used in turbo decoding: Soft Output Viterbi Algorithms (SOVA) and Maximum A Posteriori (MAP) algorithms. The MAP algorithm is more efficient but more complex than the SOVA. However, simplified versions of this algorithm such as MAX-Log-MAP and Log-MAP perform almost as well with a reduced complexity.
\\In our work, we make the choice to deal with MAX-Log-MAP decoder as it is the less complex one. The component encoders are recursive systematic convolutional encoders with generator polynomials $(5,3)$. The chosen interleaver is a "row-column" interleaver which is can simply described by Fig.~\ref{Fig:interleaver} (the data is written row-wise and read columnwise). We performed 20 iterations with data size $D$ = 10000 and block size $N$ = 1000.
\subsection{Intercept and resend attack principle}
We implement one type of eavesdropping strategy (intercept and resend) and we experiment our method on its case. Intercept and Resend is the most known eavesdropping strategy that can be implemented with actual technology means. Eve replaces Bob by applying random bases measurements to some qubits and each result will be sent to Bob without any change. After Alice and Bob public discussion about their bases measurements choices, Eve construct her one by leaving bits that correspond to Alice and Bob incompatibility measurements. We define for this eavesdropping strategy the probability $s$ that one qubit sent by Alice to Bob is eavesdropped.

\subsection{An overview of the proposed method}

\begin{figure}[h]
    \centering
    \includegraphics[width=8cm,height=1.5cm]{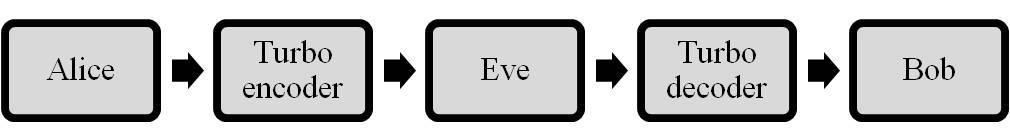}
 \caption{Proposed reconciliation method model.\label{Fig:method}}
\end{figure}

In our work, we propose to resort to turbo codes to accomplish reconciliation, more precisely the analysis model is shown in Fig.~\ref{Fig:method}. Alice and Bob communicate with BB84 protocol and Eve uses intercept and resend attack. The data sample generated by Alice is firstly encoded and than decoded in the Bob's side.

\subsection{Experimental results}

We have now collected all the ingredients to describe the results by studying secure information which is an important parameter defined by~\cite{iyed}:
\begin{equation*}
\ I_{s}=I_{AB}-I_{AE} ,
\end{equation*}

 where $I_{AB}$ is amount of information exchanged between Alice and Bob and $I_{AE}$ is the one exchanged between Alice and Eve.
\begin{equation*}
\ I_{AB}=log_{2}\left(2-\frac{s}{2}\right)-\frac{s}{4}log_{2}\left(\frac{4}{s}-1\right)
\end{equation*}
\begin{equation*}
\ I_{AE}=\frac{1}{2}log_{2}\left(2-\frac{s^{2}}{4}\right)+\frac{s}{4}log_{2}\left(\frac{2+s}{2-s}\right)
\end{equation*}

\begin{figure}[h]

    \centering
    \includegraphics[width=10cm,height=6cm]{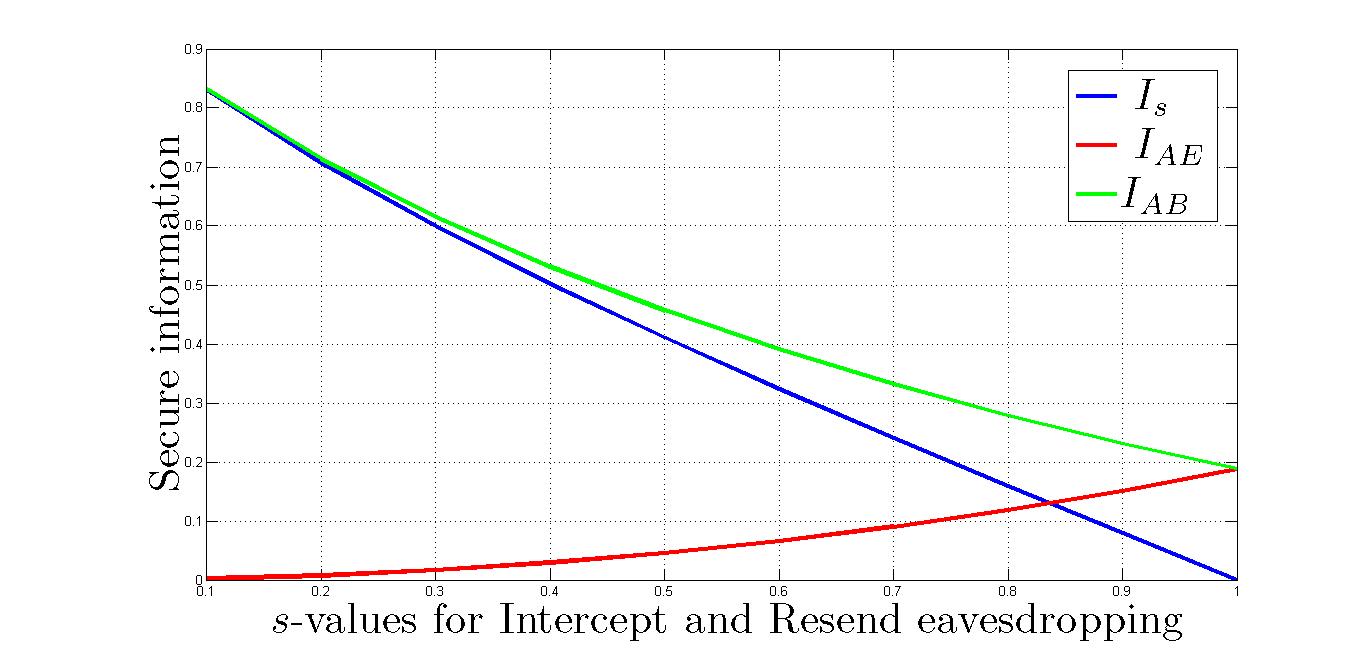}

    \caption{Secure Information as function of $s$-values for Intercept and Resend eavesdropping.\label{Fig:1}}
\end{figure}

We represent in Fig.~\ref{Fig:1} secure information $I_{s}$ as function of $s$-values for intercept and resend eavesdropping. It can be noted that secure information decreases up to 0 as the parameter of eavesdropping attack increases to 1. This means that when Eve accedes to all Alice's states, Alice and Bob will no longer have a secret and that is why the protocol should be stopped.

\begin{figure}[h]
    \centering
    \includegraphics[width=10cm,height=6cm]{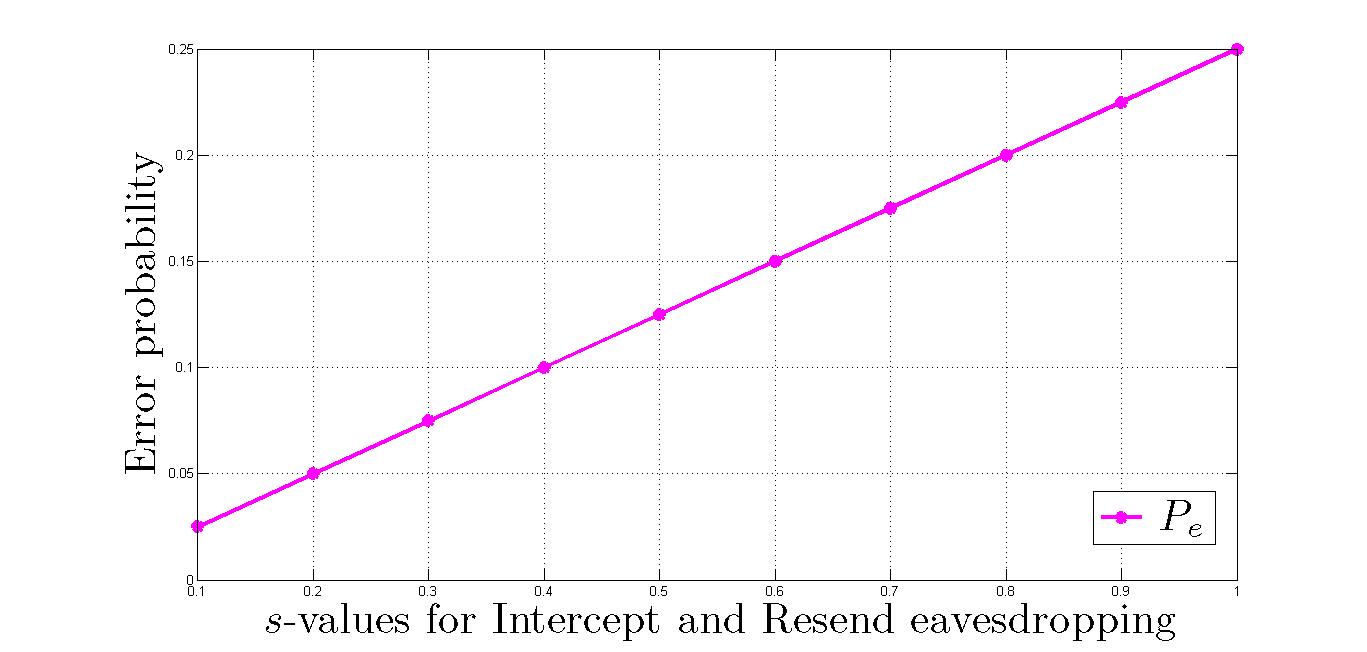}

    \caption{Error probability as function of $s$-values for Intercept and Resend eavesdropping.\label{Fig:prep}}
\end{figure}

Fig.~\ref{Fig:prep} shows error probability $P_{e}$ as function of $s$-values for Intercept and Resend eavesdropping.

 We can see clearly that $P_{e}$ is a linear function of $s$. As the eavesdropper capability to catch Alice's states increases, we have an increasing error probability.

\section{Conclusion}

Quantum cryptography allows remote parties to share secret keys. But the sifted key will contains some errors which are caused by technical imperfections, as well as possibly by Eve's intervention. Such a situation, that the legitimate partners must be removing the errors by public discussion. Our proposed method have tried to deal with this issue. \\There are many related works that are worth further investigating. First, we can generalize our research by taking into consideration not only eavesdropping effect, but also channel noise and other unwanted interactions in quantum computation and communication. Furthermore, the reconciliation of discrete random variables has been extensively studied and many practical and efficient interactive protocols (Cascade, Winnow) have been designed and are now widely used in quantum key distribution applications. However little works have been devoted to the reconciliation of continuous random variables. Thus, it seems interesting to investigate in this way. Finally, improvement could be expected by the optimization of the turbo code.

\end{document}